\documentclass[12pt,onecolumn,draftcls]{IEEEtran}
\usepackage{graphicx}
\usepackage{amsmath,amssymb,epsfig}
\usepackage{subfigure}
\usepackage[sort&compress,square,numbers]{natbib}
\usepackage{blkarray}
\usepackage{arydshln}
\usepackage{enumerate}

\newtheorem{theorem}{Theorem}
\newtheorem{proposition}[theorem]{Proposition}
\newtheorem{lemma}[theorem]{Lemma}

\newtheorem{assumption}{Assumption}
\newtheorem{claim}[theorem]{Claim}

\newtheorem{definition}{Definition}

\newcommand{\beq}{\begin{equation}}
\newcommand{\eeq}{\end{equation}}
\newcommand{\bea}{\begin{array}}
\newcommand{\ena}{\end{array}}
\newcommand{\bds}{\begin {itemize}}
\newcommand{\eds}{\end {itemize}}
\newcommand{\bdf}{\begin{definition}}
\newcommand{\blm}{\begin{lemma}}
\newcommand{\edf}{\end{definition}}
\newcommand{\elm}{\end{lemma}}
\newcommand{\bthm}{\begin{theorem}}
\newcommand{\ethm}{\end{theorem}}
\newcommand{\bprp}{\begin{prop}}
\newcommand{\eprp}{\end{prop}}
\newcommand{\bcl}{\begin{claim}}
\newcommand{\ecl}{\end{claim}}
\newcommand{\bcr}{\begin{coro}}
\newcommand{\ecr}{\end{coro}}
\newcommand{\bquest}{\begin{question}}
\newcommand{\equest}{\end{question}}

\def\bm#1{\mbox{\boldmath $#1$}}

\newcommand{\bSigma}{\mbox{$\bm \Sigma$}}

\newcommand{\avec}{{\bf{a}}}

\newcommand{\dvec}{{\bf{d}}}

\newcommand{\yvec}{{\bf{y}}}

\newcommand{\wvec}{{\bf{w}}}
\newcommand{\xvec}{{\bf{x}}}
\newcommand{\zvec}{{\bf{z}}}

\newcommand{\rvec}{{\bf{r}}}

\newcommand{\vvec}{{\bf{v}}}

\newcommand{\hvec}{{\bf{h}}}

\newcommand{\Amat}{{\bf{A}}}
\newcommand{\Bmat}{{\bf{B}}}

\newcommand{\Gmat}{{\bf{G}}}
\newcommand{\Hmat}{{\bf{H}}}

\newcommand{\Pmat}{{\bf{P}}}

\newcommand{\Tmat}{{\bf{T}}}

\newcommand{\Umat}{{\bf{U}}}
\newcommand{\Vmat}{{\bf{V}}}
\newcommand{\Wmat}{{\bf{W}}}

\newcommand{\E}{{\rm{E}}}

\newcommand{\real}{{\mathbb{R}}}
\newcommand{\comp}{{\mathbb{C}}}
\newcommand{\nat}{{\mathbb{N}}}

\newcommand{\0}{{\mathbf {0}}}

\newcommand{\define}{\stackrel{\triangle}{=}}

\def\bLambda{{\mbox{\boldmath $\Lambda$}}}

\newcommand{\be}{\begin{equation}}
\newcommand{\ee}{\end{equation}}
\newcommand{\beqna}{\begin{eqnarray}}
\newcommand{\eeqna}{\end{eqnarray}}

\usepackage{color}

\begin{document}

\title{ Spatial MAC  in MIMO Communications and its Application to Underlay Cognitive Radio}

\author{\IEEEauthorblockN{Yair Noam and Andrea J. Goldsmith \\ \thanks{This work is supported by the ONR under grant N000140910072P00006,
the AFOSR under grant FA9550-08-1-0480, and the DTRA under grant HDTRA1-08-1-0010.} }
\IEEEauthorblockA{Dept. of Electrical Engineering\\ Stanford University\\
Stanford CA 94305\\
Email: \{noamyair,andreag\}@stanford.edu} }

\maketitle
\begin{abstract}
We propose a learning technique for MIMO  secondary users (SU)  to spatially coexist with Primary Users (PU).  By learning the null space of the interference channel to the PU, the   SU  can utilize idle  degrees of freedom that otherwise would be unused by the PU.
This learning process  does not require any handshake or explicit information exchange between the PU and the SU.  The only requirement is that the PU broadcasts  a periodic beacon that is a function of its  noise  plus interference power,  through a low rate control channel.
The learning process is based on energy measurements, independent of the transmission schemes of both the PU and SU, i.e. independent of their  modulation, coding etc..
The proposed learning technique also provides a novel    spatial division multiple access mechanism for equal-priority MIMO users sharing a common channel that highly increases the  spectrum utilization compared to time based or frequency multiple access.
 \end{abstract}

\section{Introduction}

\begin{figure}
\centering
  \psfig{figure=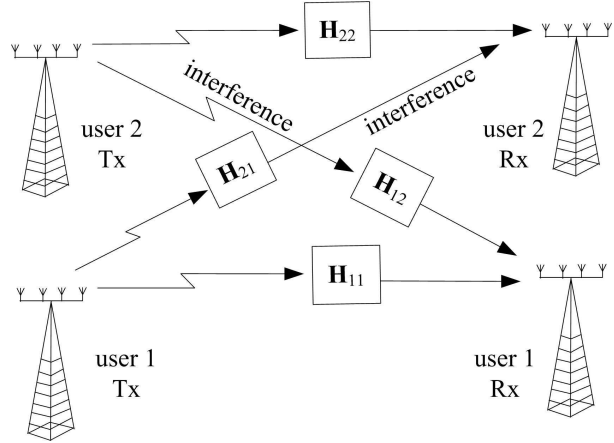,width=8cm}
\caption{ Blind spatial division multiple access for MIMO users with equal priority. The matrix   $\Hmat  _{i,j}, i\neq j\in\{1,2\} $ are  unknown to both users. The objective of the two users is to learn the null space of \(\Hmat _{ij}, i\neq j\in\{1,2\}, \).}
\label{Figure:MimoInterferenceCannel} \end{figure}

The emergence of Multiple Input Multiple Output (MIMO) communications  opens new directions  and possibilities for  spatially sharing wireless  channels \citep{JafarDegrees2007,SpencerZero-Forcing2004,RuanDynamic2011}. Consider a scenario of two independent  MIMO communication systems that share the same flat fading MIMO channel as depicted in Figure \ref{Figure:MimoInterferenceCannel}. Assuming that each user has more antennas at the transmitter then the maximum number of antennas that each one has at the receiver, they can share the channel without interfering to each other by using orthogonal spatial dimension.  This spatial sharing is even more appealing in MIMO Cognitive Radio (CR)  networks  \citep{ZhangExploiting2008,ScutariCognitive2008,ScutariMIMO2010} since it  enables a CR MIMO  Secondary User (SU)  to transmit a significant amount of  power   simultaneously as the  PU  without interfering with him by utilizes spatial dimensions  that are not used by the PU. This spatial separation requires, in both CR and MAC, that the  interference channel  be known. In the MAC (see Fig. \ref{Figure:MimoInterferenceCannel}),   it means that  \(\Hmat_{21}\) and \( \Hmat_{12}\) be known  to user 1 and 2 respectively, while in the CR case it is sufficient that the SU, say user 2, knows \(\Hmat_{21}\).  This information can be achieved by conventional channel estimation techniques that require a high level of cooperation, including handshake, transition of a known synchronous training sequence and the use of matched filters for each receiver antenna. In the MAC scenario, this process needs to be  applied twice were at the first time one of the systems transmits a training sequence while the second estimates the channel and transmits the estimation back to the other system, then it is repeated where the two systems exchange roles.  During these processes, each system must stop its data flow unless it is capable of full duplex, i.e. transmitting and receiving simultaneously at the same time and on the same frequency band.
 Although complicated, this channel estimation can be carried out in MAC since both users are equal priority. In CR on the other hand, this is far more complicated since nobody expect PUs to stop their reception and perform channel estimation for unlicensed secondary users.    Thus, acquiring  and/or  operating without knowing the interference matrix to the PU is a major issue of active research
\citep{HuangDecentralized2011,ZhangRobust2009,ZhangOptimal2011,ChenInterference2011,YiNullspace}.  in CR. Note that every solution that is good for CR problem can be utilized to the MAC problem.  Henceforth, we consider the problem of interference channel learning in the context of CR.

  We consider the underlay CR paradigm
\citep{GoldsmithBraking2009}, that is, the   SUs   are    constrained not to exceed a maximum interference level at  the PU.
The optimal power allocation for   the  case of a single SU who  knows the  matrix
 $\Hmat _{21} $
 in addition to its own Channel State Information (CSI)  was derived by Zhang and Liang
 \citep{ZhangExploiting2008}.
 In the case of multiple SUs, Scutari at al.
 \citep{ScutariMIMO2010}
formulated a  competitive game between the secondary users. Assuming that the interference matrix to the PU is known by each SU, they derived conditions  for the existence and uniqueness of a Nash Equilibrium point to the game. Zhang et al.
 \citep{ZhangRobust2009}
 were the first to take into
  consideration the fact that the  interference matrix
$\Hmat _{12} $
may not be  perfectly known  (but is partially known)  to the SU. They  proposed Robust Beamforming  to assure compliance with the interference constraint of the PU while maximizing the SU's
 throughput. Another work  for the case of an unknown interference channel with known probability distribution is due to
Zhang and So \citep{ZhangOptimal2011}  who optimized the SU throughput under a  constraint on the  probability that  the interference to the PU  be above a given threshold.

A very appealing solution concept  for CR in general and MIMO CR in particular,  is  that the SU would be able to mitigate the interference to the PU  blindly without a handshake and without using conventional channel estimation techniques. Yi
\citep{YiNullspace}
Proposed such a solution  in  the   case where there is  a channel
   reciprocity between the PU's
    transmitter  and receiver in which the SU listens to the PU signal and estimates
 $\Hmat _{12} $'s
null space  from its  second order statistics. This work was enhanced by Chen et al. in
\citep{ChenInterference2011}.
 Both works  require channel reciprocity and therefore are restricted to a PU that uses Time Division Duplexing (TDD) .
Once the SU obtains the null space of \(\Hmat _{12} \)  it can transmit within this null space without
 interfering with the PU.

 Beside the channel reciprocity case, obtaining the value of $\Hmat _{1j} $ by the SUs (i.e. the  interference channel to the PU)  requires  the PU to participate in  the SU's
  estimation  task.  This task requires that the   SU transmits  a
 training  sequence, from which the PU  estimates
 $\Hmat  _{1j} $
 and feeds it back to the  SU.  Such cooperation     increases system complexity overhead, since it
 requires  a handshake between both systems  and in addition,   the PU needs to  be synchronized to   the SU's training sequence.
This required cooperation   is one of the major  obstacles to deployment of MIMO CR systems.

\begin{figure}
\centering
  \psfig{figure=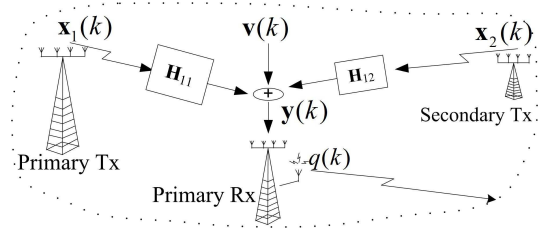,width=7cm}
\caption{ The  addressed cognitive radio scheme. The matrix   $\Hmat  _{12} $ is unknown to the secondary transmitter and $\vvec_{1}(k)$   is a stationary noise   (which may include  stationary interferences) .   The  interference from the SU,
 ${\bf H}  _{12} \xvec_2(k) $,
 is  treated  by the PU as noise, i.e. no interfere cancellation is performed.     The SU obtains a closed form expression for the  null space of the   interference channel  to the PU
$\Hmat _{12} $
 by measuring the variation of
 $q(k) $
  resulting of finite set of transmitted signals
 $\{\xvec_2(k) \}  _{t=1} ^{T} $.}
\label{Figure:SystemSetupBeacon} \end{figure}

The  objective of this paper  is to design a simple procedure based on minimal cooperation by  the PU such  that a MIMO   SU will be able to  meet its interference constraint  without explicitly estimating the matrix
$\Hmat _{1j} $
 and without burdening the PU with any handshake, estimation and/or synchronization  associated with SUs.
Consider the problem depicted in  Fig.    \ref{Figure:SystemSetupBeacon}. 
In this scheme the PU, although active, is not necessarily  aware  of the SU. Its  role  in the SU's learning process   is limited to    broadcasting    a single  one-dimensional beacon through a low rate control channel. This beacon is a  function of the PU's noise plus   interference. The advantage of this  technique over conventional channel estimation
  techniques  is that it does not require a
  handshake and synchronization
   between the
secondary and the PUs and can be implemented using only energy
measurements.
This is also  a very appealing
 property for   interference
  mitigation between two  MIMO
  users (i.e. ``multiple access")   since  it makes the
information exchange mechanism
 between the  two users   that is needed for them to share the same channel very simple.

The remainder of this paper is organized as follows:
Section \ref{Section:ProblemFormulation}  formulates the problem. Section \ref{Section:ClosedForm}  presents the Energy Based Cannel  Learning (EBCL)  algorithm for interference mitigation in  the  primary -secondary user CR scenario. Section
\ref{EqualHierarchySection}   discusses the implementation of the EBSL algorithm   in spatial channel sharing between two independent MIMO   users of equal priority.
Section \ref{Section:NumericalExamples} presets numerical results.

\section{ Problem Formulation}
\label{Section:ProblemFormulation}
Consider a flat fading MIMO
 interference channel  with  a single PU and a single  SU without
 interference  cancellation, i.e. each system treats the other
 system's signal as noise. User's \(i\)'s \(i\in \{1,2\} \)
received signal is given by
\beq
\yvec_i(k)  = \Hmat _{ii}  \xvec _i(k)  +
 \Hmat _{ij} \xvec _{j} (k)  +\vvec _i(k) ,\; k\in {\mathbb N}
\eeq
  where \(j\in\{1,2\} \), \(j\neq i\), \(\Hmat_{iq}\in\comp^{r_{i}\times t_{q}}\) and $\vvec_i(k) $ is a zero mean stationary noise. In this paper all vectors are column vectors. Let $\Amat$ be an $l\times v$ complex matrix, then, its null space is  defined as  $ {\cal N} (\Amat) =\{ \yvec\in \mathbb C^{v} :\Amat \yvec =\0 \} $ where $\0=[0,...,0]^{\rm T} $ and its column space \({\cal C}(\Amat)={\rm span}(\avec_{1},\ldots,\avec_{v})\subseteq \comp^{l}\).
  We assume that user  1 is the PU. The secondary user (user 2)   is allowed to transmit as long as the interference  does not exceed   a maximum  level at   the PU, i.e.
\beq
\label{InterferenceConstraint}
\|\Hmat _{12} \xvec _{2} (k) \| ^{ 2} \leq \eta,
\eeq
where  $\eta=0$ represents the case where the SUs are allowed to transmit only in the null space of the matrix $\Hmat _{12} $.

  Since the secondary  user is MIMO it  can share the channel without interfering with  the PU if it uses  spatially orthogonal degrees of freedom. In particular,  the SU will not interfere with the  PU  if its transmitted signal  \(\xvec _{2} \)  satisfies      $\xvec _{2}  \in{\cal N} (\Hmat _{12} )$.
 The main obstacle  in using this technique is that it requires knowledge of ${\cal N}  (\Hmat _{12} ) $.  The matrix \( \Hmat _{11} \)  is known only to the PU, and  the matrix \(\Hmat _{12} \)   is unknown to both the PU and the SU; hence its estimation   requires cooperation between the two users. The state of the art in MIMO channel estimation techniques  requires that the SU   transmits a training signal  that is known to  the PU.  The PU then  estimates the  channel   using   a matched filter. Other techniques that are not based on a known deterministic signal waveform are the blind MIMO channel estimation techniques
 \citep[e.g.]{ShengliSubspace2002,DingBlind2003,ShinBlinde2007}  in which the receiver uses the received signal  statistics, i.e covariance matrices and higher order comulant tensors,
to estimate the channel. These approaches require an extensive set of measurements  and processing at the  receiver side (the PU's receiver in this case). After the PU obtains an  estimate of \(\Hmat_{12}\) he  transmits it   to  the SU.  This kind of ``service" provided by the PU to the SU is highly undesirable due to the overhead and cooperation required on the part of the PU. Thus,  reducing the role of the PU in this channel learning phase will make CR technology more attractive for practical applications.

Our objective is to derive a simple procedure for the SU  to learn the null space of the matrix \(\Hmat _{12} \)   such that the PU would not need matched filters or to make  extra measurements other  than those required for its usual operation. We would also like to reduce the amount of processing at the PU and above all we would like the   SU  to  obtain the null space of \(\Hmat _{12}\)  without having a handshake with the PU and even without the PU   being  aware of the SU. We denote    \beq \label{DefindGmat} \Gmat\define\Hmat _{ 12}^{*} \Hmat _{12}\in \comp^{t_{2}\times t_{2}}  \eeq  and divide time into \(N\)-length intervals referred to as transmission cycles. In  each transmission cycle, the SU  transmits a constant signal (this is required only during the learning process) ,  i.e. \beq
 \label{SecondarySignal}  \begin{array} {lll}
 \xvec _{2} ((n-1) N)  =\xvec _{2}  ((n-1) N+1)  \\~~~~~~~~~~~~ = \cdots= \xvec_2 (Nn-1 ) \triangleq \tilde \xvec _{2} (n)
\end{array} \eeq
while the PU   measures its total noise plus interference. It then  broadcasts    to all of the users in its vicinity the  one dimensional signal
\(q(n) \)  that satisfies the  following assumption.
\begin{assumption}
\label{Assumption:ExtractingInterference}
There exist some \(K\in{\mathbb N} \)  such that  the value of $\Vert \Hmat _{12} \tilde \xvec _{2} (n) \Vert^{2} $  can be extracted up to an  arbitrary    scalar factor $\alpha>0$ from $\{q(l) \}  _{l=0} ^{n} $, for  every $1\leq n\leq K.$
\end{assumption}

\begin{figure}
\centering
{ \psfig{figure=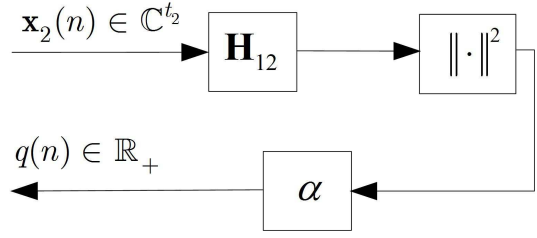,width=7cm} }
\caption{Block Diagram of the energy based cannel learning problem. The SU  objective is to learn the null space of \(\Hmat  _{12} $ by inserting a series of $\{\tilde \xvec _{2}  (n) \}  _{n\in {\mathbb N} } $  and measuring  the output \(q(n) \).  The only information that can be extracted by the SU  is that $\Vert   \Hmat _{12}  \tilde\xvec _{2} (n ) \Vert^{2}  \geq \Vert  \Hmat _{12}  \tilde
\xvec _{2} (l ) \Vert^{2} $ if $q(n)  \geq q(l) $ for every $(k-1) K \leq l,n\leq kK$ where $k\in {\mathbb N} $.}
\label{Figure:BlindLearning}
\end{figure}

Note that from the SU point of view, knowing \(\Hmat_{12}\) at  transmitter is equivalent to knowing \(\Gmat\), which is defined in \eqref{DefindGmat}.   The problem of learning the \(\Gmat\) from \(\{q(l) \}  _{l=0} ^{n}\), referred to as  the energy based cannel learning problem,  is depicted in Figure
\ref{Figure:BlindLearning}. Note that  as long as  \(\alpha \)  is constant for  every $1\leq n\leq K$,  the function \(q(n) \)  can be measured via energy detectors since \(\alpha\)  is arbitrary.

  A natural choice for a beacon that satisfies     Assumption
 \ref{Assumption:ExtractingInterference}  is  the following:
   \begin{equation} \label{ideal beacon} q(n) = \frac{1} {N} \sum _{k=(n-1 ) N+1} ^{Nn}
\E\left\{\left\Vert \yvec _{1} (k)-
 \Hmat _{11} \hat\xvec _{1} (k)\right \Vert^{2} \right\}
 \end{equation}  where   $\hat \xvec _{1} (k) $  is the decoded signal. This beacon is  transmitted at time instances $k=nN$, \(n\in\nat\). If we  neglects the  decoding errors, (i.e.  $\hat \xvec _{1} (k) =
\xvec _{1} (k) ) $  we obtain\beq\begin{array} {lll}
q(n) =\frac{1} {N} \sum _{t=(n-1) N} ^{Nn-1}
\E\left\{\Vert\Hmat _{12}  \tilde \xvec _{2} (n)  +\vvec(k) \Vert^{2}  \right)
\\~~~~~=\Vert \Hmat  _{12} \tilde \xvec _{2} (n) \Vert^{2} +{\rm Tr} (\E\{
\vvec _{1} (k) \vvec _{1} ^*(k) \} )  \\~~~~~= \tilde \xvec _{2} ^*(n)  \Gmat \tilde \xvec _{2} (n) +c
\end{array} \eeq

We will now show that this beacon satisfies Assumption
 \ref{Assumption:ExtractingInterference}, i.e. that the secondary user can extract \( \alpha\Vert\Hmat _{12} \tilde \xvec _{2} (n)  \Vert^{2 } \)  from \(\{q(l) \}  _{l=0} ^{n} \). This is done as follows:
  At the beginning of the learning process (\(n=0) \)  the  SU transmits \(\tilde \xvec _{2}(0)=0\), that is, it does not transmit. Let \(\alpha>0\)  be the magnitude of the control channel from the PU to the SU. Then, at time  \(k=0\)  the SU measures \(\alpha q(0) \)     where  \(q(0) ={\rm Tr} (\E\{\vvec _{1} (k) \vvec _{1} ^* (k) \}\).    For \(n> 0\), the SU   transmits  the  signal  $\tilde\xvec _{2} (n) $    and at time $k=nN$ it measures the  $\alpha q(n) $  broadcast by the PU. The SU   then  obtains $ \alpha\Vert \Hmat _{12} \tilde \xvec _{2} (n)  \Vert^{2} $   by subtracting $\alpha q(0) $ from $\alpha q(n) $.
Note that \(\alpha\)   may be unknown to the SU and that  the only requirement is that it  be constant during the learning process.

In practice, the beacon will be based on the sample average
 \begin{equation} \label{beacon} q(n)  =
 \frac{1} {N} \sum _{k=(n-1) N} ^{Nn-1}
\left\Vert
 \yvec _{1} (k) - \Hmat _{11} \hat \xvec _{1} (k)  \right\Vert^{2}
  \end{equation}
 which depends on the  averaged   value of $ \Vert \zvec(k) \Vert$  at the $n$th cycle where  \beq \label{DefineZ}\zvec(k) =\Hmat _{12}  \xvec _{2} (k)  +\vvec _{1} (k) \eeq It is important to stress  that   the function \(q(n) \)  is calculated entirely from \(\yvec _{1} (k) \). Therefore it  is  calculated by the PU  processing  unit after decoding its signal  \(\hat \xvec _{1}  (k) \)  without any additional measurements.

 In the next section we will show how the SU can learn the null space of the matrix $  \Hmat  _{12} $ from the measurements \(\{q(n) \}   _{n=1} ^{t_{2}} \), where \(t_{2}\)  is the SU's number of transmit antennas.

\section{The energy based cannel  learning algorithm}
 \label{Section:ClosedForm}
In order to obtain \( \Hmat _{12} \)'s null space it is sufficient to calculate \(\Gmat\)'s null space (where \(\Gmat\) is defined in \eqref{DefindGmat}). The following proposition expresses the matrix  $\Gmat$ as a function of $\{\tilde \xvec _{2} (n) \Gmat\tilde \xvec _{
2} (n) \}  _{n=1} ^{t_{2} ^{2} } $, where each $\tilde \xvec _{2} (n) $ is a different transmitted signal.
 \begin{proposition}
 \label{StrightForwardApproach}  Let \(S(
\Amat,\xvec) \define\xvec^{*} \Amat \xvec\) and  $\rvec _{l,m} (\theta,\phi) $ be  a \(t_{2}-  \)dimensional  column vector whose entries are all equal to zero except of the \(l\)th  and \(m\)th entry, which are equal to \(\cos(
 \theta) \)  and \(e^{-i\phi} \sin( \theta) \), respectively, i.e.
\beq
\begin{array} {lll}
\rvec _{l,m} (\theta,\phi) =[0, \cdots ,0, \cos(
\theta) ,\\\;\;\;\;\;\;\; \;\;\;\;\; \;\;\;
\;\;0, \cdots, 0,e^{-i\phi }  \sin( \theta) ,0, \cdots,0]^{\rm T} \end{array} \eeq
 The entries \(\{g _{l,m} \}  _{l,m=1} ^{t_{2}  } \)  of the  matrix $\Gmat= \Hmat_{12}^*\Hmat_{12}$ are given by
\beqna \label{gll} g _{l,l} &=& S \left(\Gmat, \rvec _{l,m} (0,0)  \right)  \\\label{Real_glm}
\Re (g _{l,m} ) &=&c _{l,m} (\pi/4,0)  \\
\label{Imaginary_glm}
\Im (g _{l,m} ) &=&-c _{l,m} (\pi/4, \pi/2)
\eeqna
where
\beq \label{define clm} \begin{array} {lll}
c _{l,m} (\theta,\phi)  &=(g _{l,l} \cos^{2}  (\theta ) + g _{m,m}  \sin ^{2} (\theta) )  \\&~~~-S\left(
\Gmat,\rvec _{l,m} (\theta,\phi)  \right) \\&
\end{array} \eeq
\end{proposition}
\IEEEproof
Note that \beq
\label{TrigonometricS} \begin{array} {lll}
S(\Gmat,\rvec _{l,m} (\theta,\phi)  ) =\cos ^2(\theta )  \left|g _{l,l} \right| \sin ^2(\theta )  \left|
g _{m,m} \right|\\~~~~~~~~~~~ ~~~ ~~~~  ~~~~~ -\left|
g _{l,m} \right| \sin (2 \theta )  \cos ( \phi+\angle g _{l,m}  )
\end{array} \eeq
from which   \eqref{gll}   follows. By substituting \eqref{TrigonometricS}  into \eqref{define clm}  we obtain   \beq\label{sinus}
c _{l,m} (\theta,\phi) =\sin (2 \theta )  \left|g _{l,m} \right| \cos \left( \phi +\angle g _{l,m} \right)
\eeq
from which   \eqref{Real_glm}  and  \eqref{Imaginary_glm}    follow.
  \hfill $\Box$

The  EBCL algorithm provides a closed form expression for the matrix $\Gmat$.  For every $\tilde \xvec$, the value of $\Vert \Hmat\tilde\xvec \Vert^{2} $ can be obtained by transmitting $\tilde \xvec(n) $, receiving $q(n) $ and subtracting $q(0) $ from it, i.e. \beq \Vert \Hmat\tilde\xvec(n) \Vert^{2}  =q(n) -q(0) \eeq
From Proposition \ref{StrightForwardApproach}, it follows that    the matrix $\Gmat$ can be obtained precisely by \(t_{2}  ^{2} \)  transmission cycles.
The CF-BNSL algorithm is described in Table \ref{Table:CF-BNSL Algorithm}.
\begin{table}
\caption{The EBCL algorithm}
\label{Table:CF-BNSL Algorithm}
 function $\Gmat$={\bf EBCL}
\begin{description}
\item Set $b={\rm S} (\Gmat,\0) $;
\item for $l=2,...,t  $
\begin{description}
\item Set $g _{ll} ={\rm S} (\Gmat ,\rvec _{l,l} (0,0) ) -b$;
\end{description}
\item end for
\end{description}
\begin{description}
\item for $l=1,\dots, t  -1$
\begin{description}
\item for $m=l+1,...,t  $
\begin{description}
\item Set $\alpha _{l,m } ={\rm S} (\Gmat ,\rvec _{l,m} (\pi/4,0) -b$; \item Set $\beta _{l,m} ={\rm S} (\Gmat ,\rvec _{l,m} (\pi/4,\pi/2) ) -b$;
\item Set $c _{1} ={\rm Calc\underline~c} (g _{ll} , g _{mm} , \alpha _{l} ,\pi/4) $;
\item Set $c _{2} ={\rm Calc\underline~c} (g _{ll} , g _{mm} , \beta _{l} ,\pi/4) $;
\end{description}
end for
\end{description}
end for
\end{description}
end {\bf EBCL} \\
 function: $c=${\bf Calc\underline~c} $(g_1,g_2, \alpha,\theta) $
\begin{description}  \item $c=g _{1} \cos^{2} (\theta) +g _{2} \sin^{2} (\theta) -\alpha; $
\end{description}
 end {\bf Calc\underline~c}
\end{table}
After obtaining the matrix $\Gmat$, its   null space  can be calculated offline at the secondary transmitter's processing unit.
Once the  SU knows the null space of the  interference channel to the PU's transmitter it can transmit freely as long as its transmitted  signal is restricted to its null space, i.e. \(\xvec _{2} \in {\cal N} (\Hmat _{12} )\).

The advantage of the proposed scheme (see Fig. \ref{Figure:SystemSetupBeacon})  is that the  PU, although active, does not  have to be aware of the SU. Its  role  in the SU's learning process   is limited to    broadcasting    periodically the  beacon \(q(n) \)  through a low rate control channel to all of the secondary users in its vicinity. Thus, in order to implement the EBCL algorithm,  the secondary user needs only to detect and measure    \(q(n)\)'s energy in every transmission cycle  without having a handshake with the PU.  Recall that the  only condition  required for the EBCL  is that Assumption
 \ref{Assumption:ExtractingInterference}  holds. This assumption holds even if there  are multiple secondary users in the system as long as their interference to the PU is stationary. Thus a new secondary user can join the network while  multiple SUs coexist with the PU in a steady state, i.e. they are not varying the spatial orientation or their transmitted signal.

 \section{ EBCL algorithm for spatial division multiple access }  \label{EqualHierarchySection}

  The fact that the CF-BNSL algorithm  is based entirely on energy  measurement and not on  matched filters makes it  very appealing for implementation  as a blind spatial division multiple access technique for MIMO users with equal priority (see Figure 3), that is,  \eqref{InterferenceConstraint}   is no longer required.
This simplifies the  coordination between the two users  as follows: At the first stage, there is a handshake between the two systems in which it is decided which system begins with  learning and which provides feedback. Assume that system  2 begins with  learning while system No. 1 feeds back its measurements. Then  system No. 2 transmits a signal  \(\tilde\xvec _{2} (n)   \)  while system No. 1 measures and feeds back the beacon in \eqref{beacon}. This way, system \(2\)  learns the matrix
\(\Gmat _{1} \)  by applying the CF-BNSL algorithm. This process is then repeated where  both systems exchange roles such that  system 1 learns \(\Gmat _{2} \).
Thus, if  system  1 and 2  restrict their transmission to \({\cal N} (\Hmat _{21} ) \)  and \({\cal N} (\Hmat _{12} ) \)  respectively,   they do not interfere with each other and create in effect a Spatial Channel Sharing (SCS) .

An important question that  arises is whether the spatial channel sharing is worth the effort of null space  learning. Recall that in the primary-secondary user CR scenario the SU must be invisible to the PU. This fact makes the learning of  \({\cal N} (\Hmat _{12} ) \)  worthwhile because,  as long as the channel remains unchanged, the SU is operating freely without colliding.   This is not the case for  MIMO users of equal priority. They can choose not to mitigate interference at all or to    share the channel using a much  simpler multiple access scheme such as Frequency Division Duplexing (FDD),  which is static and   does not require null space learning.
 In the sequel it is shown that the SCS provides a much better spectrum utilization (in terms of degrees of freedom) than FDD if both  systems have a sufficient number of antennas at the transmitter.

In the sequel it is assumed that  \(  1\leq i\neq j\leq 2\),  \( {t_{i}}> {r_{j}}\) and  that the EBCL algorithm is performed by both users. Let \beq\label{DefineGi}\Gmat_{i}=\Hmat_{ji}^{*}\Hmat_{ji}\eeq and let \beq\label{DefineWi}\Wmat_{i}\bLambda_{i}
\Wmat_{i}^{*}=\Gmat_{i}\eeq be its eigenvalue decomposition. Then  user \(i\)'s  pre-coding matrix \(\Tmat_{i}\) is given by
\beq
\label{DfineT}
\Tmat_{i}=[\wvec_{q_{1}},...,
\wvec_{q_{{t_{i}}-{r_{j}}}}]
\eeq
 where \(\wvec^{k}_{q}\) is \(\Wmat_{i}\)'s \(q\)th column and  \(q_{1},q_{2},...,q_{ {t_{i}}
- {r_{j}}}\) are the  indexes that chose the eigenvectors that correspond to \(\Gmat_{i}\)'s Null space, i.e.
\beq
\wvec_{q_{1}}^{*}\Gmat_{i} \wvec_{q_{1}} =\cdots =\wvec_{{t_{i}}-{r_{j}}}^{*}\Gmat_{i}
\wvec_{{t_{i}}-{r_{j}}}=0\eeq

The following proposition shows that  for Zero-Mean Spatially White\footnote{It means the the entries of the matrix \(\Hmat\) are i.i.d. zero-mean unit-variance circular  Gaussian  random variables \citep[see e.g.][Section 10.1]{goldsmith2005wc}.} (ZMSW) channels that satisfy \(t_{i}\geq r_{j}\), the EBCL results in a free interference   \(r_{i}\times(t_{i}-r_{j})\)-ZMSW channel for each user.
\begin{proposition} \label{LargeAntennaProposition} Assume that \(\Hmat_{iq}, q,i\in\{1,2\}\)  are      \({r_{i}}\times {t_{q}}\)  (ZMSW) channels  that are independent of each other and satisfy
\( {t_{i}}\geq {r_{j}}\).   Let \(\tilde \Hmat_{ii}\) be user \(i\)'s equivalent channel when  both users apply  the CF-BNDL algorithm i.e.    \(
\tilde \Hmat_{ii}=\Hmat _{ii} \Tmat _{i} \) where  \(\Tmat  _{i} \)  is users \(i\)'s pre-coding matrix defined in \eqref{DfineT}. Then, \(\tilde \Hmat_{ii}\) is an \( {r_{i}}\times ( {t_{i}}- {r_{j}})\) ZMSW channel.
  \end{proposition}
\IEEEproof See Appendix
\ref{Appendix:ProofOfLargAntennaProposition}.

Proposition \ref{LargeAntennaProposition} implies that if  \(t_{i}\geq r_{i}+r_{j}\) for \(i\neq j\in\{1,2\}\), the difference between  SCS using the EBCL algorithm compared to the case where there is no interference is equivalent to not using \(r_{j}\) antennas.   Furthermore,  both users would not lose degrees of freedom compared to the case where there is no interference  since \( {\rm rank}(\Hmat _{ii})=r_{i}\; { a.s.}, \text{and } {\rm rank}(\tilde\Hmat_{ii})=\min\{r_{i},t_{i}-r_{j}\}\; a.s.\)  which  are equal    if \(t_{i}-r_{j}\geq r_{i}\). The following theorem extends the last  statement for a wider range of channel types.

\begin{theorem} \label{PropositionNoLoss}  Assume that \(\Hmat _{iq} \) \(i,q\in\{1,2\}\) are    independent (i.e. independent of each other) random matrices defined on the same   probability space \((\Omega,{\cal F} ,P) \) such  that \({\rm vec }(\Hmat _{ii})\) is a continues random vector\footnote{A \( {t}\)-dimensional complex  random vector  \(\xvec\) is said to be continuous if it can be written as \(\xvec=\xvec_{\rm Re}+i\xvec_{\rm Im}\) where  $\tilde \xvec=[\xvec_{\rm Re}^{\rm T},\xvec_{\rm Im}^{\rm T}]$ such that \(\tilde \xvec\) is a continuous $2 t$-dimensional random vector, i.e. \(\tilde \xvec\) has a probability density function with respect to  the  Lebesgue measure.} for \(i=1,2\). Let \(\dvec _{i} ={\rm rank } (\Hmat _{ii} )  \)  be user \(i\)'s number of degrees of freedoms if he is operating alone, and let \(\dvec^{n}  _{i} \)  be user \(i\)'s number of degrees of freedom when both users apply  the EBCL algorithm, i.e. \(\dvec _{i} ^{n} =  \text{rank} (\tilde\Hmat_{ii}) \), where \(
\tilde \Hmat_{ii}=\Hmat _{ii} \Tmat _{i} \) and \(\Tmat  _{i} \)  is users \(i\) 's pre-coding matrix defined in \eqref{DfineT}. Then, \(\dvec _{i} =\dvec  _{i} ^{n} \; a.s. \)  if \({t  _{i} } \geq {r  _{i} } +{r  _{j} } \).
\end{theorem}

 \IEEEproof See Appendix
 \ref{Appendix:ProofOfProposition}.

\section{Obtaining Additional Degrees of Freedom}\label{Section:NumericalExamples}

\begin{figure}
\centering
  \psfig{figure=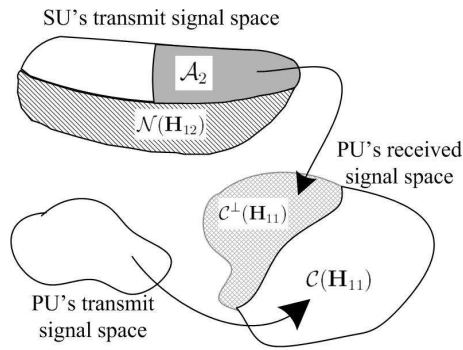,width=6cm}
\caption{Graphical illustration of the space \({\cal A}_{2}\). Assuming that all matrices  are full rank and that the secondary user has more antennas ate the receiver than at the transmitter, i.e. \(r_1>t_{1}\), then \({\cal  C}^{\bot}(\Hmat_{11})\neq 0\). Then, the  subspace \({\cal A}_{2}\)  that maps  signals to \({\cal C}^{\bot}(\Hmat_{11})\) can be used by the SU without interfering with the PU. A necessary and sufficient  condition is that  \(t_{2}>t_{1}\).}
\label{Figure:IllustrationSetA} \end{figure}

Constraining the SU to  transmit only in ${\cal N}(\Hmat_{12})$ may be inefficient in some cases. Consider a scenario where the PU has more antennas at the receiver than at the transmitter i.e.   ${t_{1}}
<{r_{1}}={\rm rank }(\Hmat_{11})$  and full CSIR of its own  channel $\Hmat_{11}$. Then, the PU's   signal of interest at the receiver, that is  \(\Hmat_{11}\xvec_{1}\), can  lie only  in  the
 ${r_{1}}$-dimensional subspace  ${\cal C}
(\Hmat_{11})\subset{\mathbb C}^{{r_{1}}}$. This redundancy   can be
utilized
by the SU  to obtain additional degrees of freedom by transmitting         $\xvec _{2}  \in{\cal N} (\Hmat _{12} ) + {\cal A}  _{2} $\footnote{The sum of two vector subspaces is the vector space created by the sum of all the vectors in these two subspaces, i.e. let \(\cal B\) be a vector space and let \({\cal B}_1,{\cal B}_2 \) be two vector subspaces of \({\cal B}\), then \({\cal
 B}_1+
{\cal B}_2=\{\xvec\in{\cal B}:\xvec=\yvec
+\zvec, \yvec\in{\cal B}_1,\zvec={\cal} \in {\cal B}_2\}.\) } where  ${\cal A}  _{2} =\{ \xvec _{2} \in {\mathbb C} ^{t _2
 }:{\Hmat  _{12}  \xvec \in \cal C} ^{
\bot}  (\Hmat _{11} ) \}$ (see Fig.
 \ref{Figure:IllustrationSetA} for illustration). If all matrices are full rank, a necessary and sufficient condition for  ${\cal N}(\Hmat _{12} ) + {\cal A}  _{2}\neq 0$ is that \({r_{2}}>r_{1}\).   Note that the subspace ${\cal N} (\Hmat _{ 12} ) + {\cal A}  _{2} $ is equal to ${\cal N} (\Pmat _{ \Hmat_{11} } \Hmat _{12} ) $ where \(\Pmat _{ \Bmat} = \Bmat ^{*} (\Bmat \Bmat^{*} ) ^{\dagger} \Bmat \)  is the projection matrix into the column space of \(\Bmat\)  (which is equal the rang  of $\Bmat$) and \( (\cdot) ^{\dagger} \)
  represents the pseudo inverse  operation.
 These extra degrees of freedom can be obtained  by the  EBCL algorithm with no additional cost. The only modification required is for the PU to project $\zvec(t)$, defined in \eqref{DefineZ},  into ${\cal C}(\Hmat_{11})$ while the rest of the algorithm remains the same, i.e. to replace  $ \zvec(t)=\yvec_{1}(t)-\hat
\xvec_{1}(t)$ with  $\tilde\zvec(t)=\Pmat_{\Hmat _{11}}\zvec(t)$.
    This idea can also be implemented in the case of two  users with equal priority that is described in Section \ref{EqualHierarchySection}.

\section{Numerical Examples}

\begin{figure}
\centering
  \subfigure[ ]{\label{SubFigure:SpatialDivisionDuplexing}
  \psfig{figure=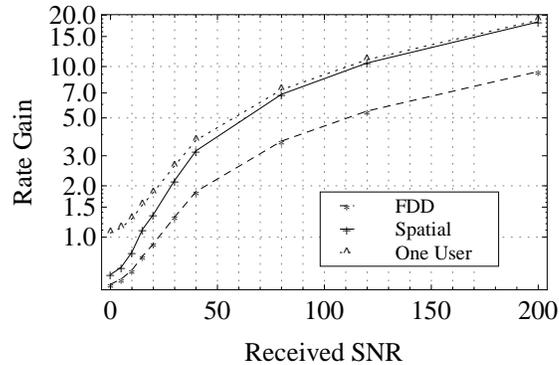,width=8cm} }
  \\ \subfigure[]{ \label{SubFigure:CapacityGainAntennaNumber}  \psfig{figure=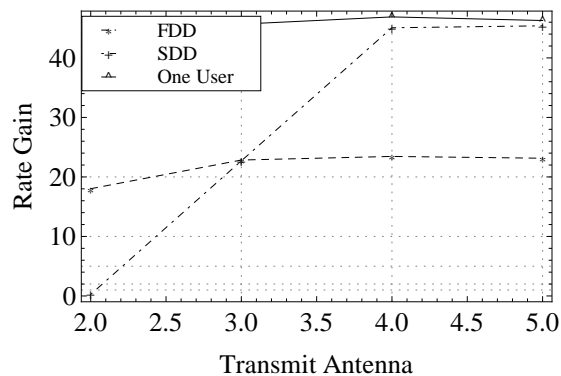,width=8cm} }
\caption{ Comparison between blind spatial division  and FDD/TDD in a symmetric MIMO interference interaction. The matrices $\Hmat   _{i,q} $ \(i,q\in \{1,2\} \)  are  i.i.d. complex Gaussian with zero mean and unit variance. Both user's  number of  received antennas is 2. The interference expected power is 10.5dB lower than the expected signal power for both users. The  vertical axis represents the ratio between the achievable rate to the rate obtained via uniform power allocation over the entire band/time. In Subfigure (a)  The horizontal axis represents expected received SNR  while the number of transmit antennas for each user  is 2. In Subfigure (b)   the horizontal axis represents the number of antennas at the transmitter while the  expected power at each receiver is  140 dB.}
\label{Figure:SpatialDivisionDuplexing}  \end{figure}

To determine the value of null space learning in this setting we turn to simulations. Figure
 \ref{Figure:SpatialDivisionDuplexing}   compares the rate gain of SCS over that of FDD in a two-user symmetric  MIMO  interference channel without interference cancellation. By symmetric we mean that \({t  _{1} } ={t  _{2} } \) ,  \({r  _{1} } ={r  _{2} } \)  and that \(\Hmat  _{ii} ,i=1,2\)  are ZMSW channels as well as \(\Hmat  _{ij} ,i\neq j,\in\{1,2\} \). Figure \ref{SubFigure:SpatialDivisionDuplexing}   shows that for \({t} =4\)  and \({r} =2\) , the  SCS outperforms the  FDD, i.e. the SCS's rate gain is higher than that of the FDD. Furthermore,  in the  high SNR regime the SCS rate   converges to the channel capacity without  interference, i.e. the rate of a single user occupying the entire channel,  as long as \({t} \geq 2{r}\), as  shown in Figure
 \ref{SubFigure:CapacityGainAntennaNumber}. From this we conclude that  in  the FDD  scheme, each user exploits only  half of its degrees of freedom, whereas  in the SCS scheme both users exploit all of their degrees of freedom (as long as \(t_{i}\geq r_{i}+r_{j}\)) and the only performance loss is due to the restriction of the transmit signal to \({\cal N}(\Hmat  _{ji} ) \).

It is important to stress that knowing   \(\Gmat\)  can be utilized for a more sophisticated  channel sharing than the SCS.
For example, suppose that in addition to transmitting in  \({\cal N}  (\Pmat  _{ \Hmat  _{jj} }  \Hmat  _{ji} ) \),    system \(i\)  wishes   to use also part of its orthogonal compliment  \({\cal N} ^{\bot} (\Pmat  _{ \Hmat  _{jj} } \Hmat  _{ji} ) \). This of course creates interference to system \(j\). However by choosing  eigenvectors  that  correspond to \(\Gmat\)'s  lowest eigenvalues, system \(i\)  can balance between its performance gain and the  interference to system \(j\). To show that explicitly, let \(\Vmat\bSigma\Vmat^{*} \)  be the eigenvalue decomposition   of \(\Gmat\), where \(\bSigma\)  is a real nonnegative diagonal matrix that contains \(\Gmat\)  eigenvalues in  decreasing order, i.e. \(\sigma  _{1}  \geq\sigma _{2} \geq\cdots\geq\sigma  _{d} >0\)  where \(d<{t  _{i} } \). Then the eigenvector that corresponds to \(\sigma  _{d} \)  (i.e. \(\Vmat\)  \(d\) 's column)  produces minimum interference to system \(j\). This way, system \(i\)  can balance between choosing eigenvectors that provide it with the best performance gain and minimizing the interference to system \(j\).
\section{conclusions}

We proposed a  blind technique for MIMO SUs to spatially coexist with PUs based on  minimal cooperation from the PU. This cooperation does not require additional sensing by the PU  and is  carried out by calculating the power of  the PU's total noise plus interference. This value is broadcast via a low rate control channel to all of the SUs in its vicinity (beacon). By doing so, the PU enables the SU to utilize unused degrees of freedom.

The advantages of the proposed technique are:
\begin{enumerate}
\item The SU operates autonomously and independently of the  PU (as long as the PU transmits the defined  beacon).
\item The PU produces the beacon from information that already exists in all communication systems, i.e. from the PU's decoded signal and its received signal.
\item The entire learning process is based on energy measurements, independent of the transmission schemes of both the PU and SU, i.e. independent of their  modulation, coding etc..  This flexibility is very important in CR networks which are inherently ad-hoc.
\item The entire learning process takes  \({t_{2}} ^{2} \)   transmission cycles where \(t_{2}  \)  is the number of the SU's transmit antennas.
\item The proposed technique is easily applicable to CR networks with one PU and multiple SUs as long as  only one SU performs  the learning procedure at a time while the other SUs don't change their spatial power allocation.
In practice, this is  not a problem since the learning process takes only \({t}_{2} ^{2} \)  transmission cycles.    \end{enumerate}

For the same reasons  the proposed scheme can be  easily implemented  for spatial channel sharing of two independent MIMO  secondary users of equal priority.  We   demonstrated that if both users share the channel using  the  CF-BNSL algorithm:
\begin{enumerate}\item They  don't loss degrees of freedom while gaining an interference free MIMO channel.
\item In case of for zero-mean spatially-withe  Gaussian channels and \(t_{i}>r_{j}\), then the SCS results in a free interference   \(r_{i}\times(t_{i}-r_{j})\)-zero-mean spatially-withe   Gaussian  channel for each user.\end{enumerate}

\appendices

\section{Proof of Proposition
 \ref{LargeAntennaProposition}}
 \label{Appendix:ProofOfLargAntennaProposition}
Without loss of generality we set \(i=1,j=2\) and denote  \(\check\Hmat_{1}=\Hmat_{11}\Wmat_{1}\).   Since \(\Hmat_{21}\)  is ZMSW channel, the random matrix \(\Gmat_{2}\) (defined in \eqref{DefineGi}), by definition, is  a central Wishart Matrix.   Thus, \(  \Wmat_{1}\) (defined in  \eqref{DefineWi}) is  a unitary matrix that is  uniformly distributed over the manifold of unitary matrices in \(\comp^{ {t_{1 }}\times  {t_{1}}}\) \citep[see e.g.,][Lemma 2.6]{tulino2004random}. Since the   channel  \(\Hmat_{11}\)  is ZMSW it is   bi-unitary invariant
  \citep{tulino2004random}, that is  \( \Umat
\Hmat_{11}\Vmat\)'s  distribution is unchanged for any unitary matrices \(\Umat,\Vmat\). Thus, for every  \(\Wmat_{1}\), the conditional distribution of
 \(\check\Hmat_{1}\)'s is equal to \(\Hmat_{11}\), i.e. \(P(\check\Hmat_{1}/\Wmat_{1})=P(\Hmat_{11})\). Therefore, given \(\Wmat_{1}\),  \(\check \Hmat_{11}\) entries are   i.i.d. zero-mean unit-variance complex Gaussian random variables (i.e. ZMSW channel) and because this distribution  is not a function of \(\Wmat_{1}\),  the  marginal distribution of \(\check \Hmat_{1}\) is the  same, i.e. \(P(\check\Hmat_{1})
=P(\check\Hmat_{1}/\Wmat_{1})\). It follows that  \(\check \Hmat_{1}\) is a \( {r_{1}}\times  {t_{1}}\) ZMSW  channel  and therefore \(\tilde \Hmat_{1}\) (which is composed of some  \( {t_{1}}- {r_{2}}\) columns of  \(\check \Hmat_{1}\)) is an \( {r_{1}}\times ( {t_{1}} - {r_{2}})\)  ZMSW channel.

\section{Proof of Proposition \ref{PropositionNoLoss}}
\label{Appendix:ProofOfProposition}
In this proof we shall use some special notation. Matrices will be denoted by italic  upper case letters (i.e. the channels \(\Hmat_{iq},i,q=1,2\) are now denoted by \(H_{iq},i,q=1,2\)) while   random matrices will be denoted by boldface upper case letters.
We will make not notational distinction  between  scalars and vectors and denote both with lower case italic letters.   Random vectors/variables will be denoted by boldface lowercase letters. Without loss of generality, we set  \(i=1\) and  denote \(H_{11}^{\rm T}=H\) and \(H_{12}^{\rm T}=\tilde H\).   Let  \(h_{q},
\tilde
h _{q} \)  be  \(H\)'s and \(\tilde H\)'s \(q\)th  columns respectively and \(H_{-q}\) be the \(  {t_{1}}\times ( {r_{1 }}-1)\) matrix that results from deleting \(H\)'s \(q\)th column.

The Theorem is  first proven   for real matrices. In this case    \(\hvec _{q}, \tilde \hvec_{q} :\Omega\longrightarrow \real^{  { t_{1} } } \)  are  Borel measurable functions.
 If   \( {r_{1} } \leq  {t_{ 1} } - {r_{ 2} } \),  user 1 losses at least one degree of freedom iff there exists a sequence of scalars   \(\{a_{q}\} _{q=1}^{ {r_{1}}} \) not all zero  such that \(\sum_{q=1}^
{ {r_{1}}}a_{q} h_{q} ^{ } \in {\cal N} ^{ \bot} (H_{ 21} )= \text{span}  ( \tilde h _{1} ,...,\tilde h_{  {r_{2} } } )\).  The later is equivalent to the following statement: There exists \( 1\leq q\leq  {r_{1}}\) such that \(\hvec_{q}\in{\cal C}(\Bmat _{-q})\)
 where   \(\Bmat_{-q}\define [\Hmat_{-q}, \tilde\Hmat]\).
 Using the sub-additivity of measures
\beq
P(\dvec_{1} ^{N} <\dvec_{1} ) \leq P\left( \bigcup_ { q= 1} ^{ {r _{1} } }  \hvec  _{q} \in {\cal C}(\Bmat_{-q})   \right) \leq \sum_{ q=1} ^{  {r _{1} } } P\left(\hvec  _{q}  \in {\cal C}(\Bmat_ {-q} ) \right)
\eeq Note that\footnote{The existence of a conditional probability measure \(P(\cdot\vert \hvec_{q})(\omega)  \) for each \(\omega\in \Omega\) is due to the fact that all random vectors are assumed to be  \(\real ^{ {t_{1}}}\)-Borel measurable. Such probability measure is termed  regular conditional probability \citep[see e.g.][]{athreya2006measure}.} \beq P(\hvec_{q}\in{\cal C}(\tilde\Bmat _{q}) )=\int_{\Omega}P(\hvec_{q}\in{\cal C}(\Bmat_{-q}) \vert\tilde\Hmat)dP(\omega) \eeq
It remains to show that \(P(\hvec_{q} \in{\cal C}(\Bmat_{q})\vert \tilde\Hmat )=0, a.s\). By hypothesis,  \(\Hmat \) is independent of \(\tilde\Hmat\), thus \(P(\hvec_{q}\in{\cal C}(\Bmat_{q})
\vert\tilde\Hmat)=P(\hvec_{q}\in{\cal C}(\Bmat_{q})), a.s\).   Now recall that \(P_{\Hmat}\) is absolutely continuous with respect to the  Lebesgue measure, that is, \( P_{\Hmat}<<m^{ {t_{ 1} } {r_{1}}}\)\footnote{Let \(\mu,\nu\)  be two measure defined on the same measurable space \((X,{\cal M} ) \) , then \(\mu << \nu\)  if \(\nu(A) = 0\Rightarrow \mu(A) =0 \). }    where \(m^{k } \)  is the \( k\)- dimensional  Lebegsue measure. Let \({\cal Q}({Z})=
 \{[x,Y]:x\in{\cal C}([Y,Z]),
Y
\in\real^{ {t_{1}}\times( {r_{1}}-1)}, Z\in\real^{ {t_{1}}\times  {r_{2}}} \}\) and let \({\cal Q}_{Y}(Z)=\{x:[x ,Y]\in {\cal Q}(Z)\}\) be \({\cal Q}(Z)\)'s \(Y\)-section. Then for every \(Z\in
\real^{ {t_{1}}\times  {r_{2}}}\)
\beq
m^{t_{1}\times r_{1}}({\cal Q}(Z))=\int_{\real^{ {t_{1}}
\times ( {r_{1}}-1)}}m^{ {t_{1} }}({\cal Q}_{Y}(Z)) dm^{ {t_{1}}\times(r_{1}-1)}(Y)
\eeq
\citep[see e.g.][Theorem 2.36]{folland} and since  for every \(Z,Y\), \({\cal Q} _{Y}(Z)\)  is a vector subspace of \(\real^{ {t_{1}}}\) whose dimension is at most \( {r_{1}}+ {r_{2}} -1\) it satisfies  \(m^{ {t_{1}}} \left({
\cal Q}_{Y}(Z ) \right)=0\) (recall that \( { r_{2} } + {r_{1}} \leq  {t_{1} } \)). This establishes the desired result for real  channel matrices.

To extend this result to complex matrices, note that    \(\hvec^{} _{q} =\hvec^{} _{q,\text{Re} } +i \hvec^{} _{q,\text{Im} } \), and \(\tilde\hvec^{} _{q} =\tilde\hvec^{} _{q,\text{Re} } +i \tilde\hvec^{} _{q,\text{Im} } \) where \(\hvec^{} _{q,\text{Re} },\hvec^{} _{q, \text{ Im}}, \tilde\hvec^{} _{q,\text{Re} } ,\tilde\hvec^{} _{q,\text{Im} } : \Omega \longrightarrow \real^{ { t_{ 1} } } \)  are Borel measurable functions. Furthermore,    the vector space \(\comp^{ {t_{1}}} \)   is isomorphic to \(\real^{2 {t_{1}}}\), that is, there exists a bijective mapping  (one to one and on to) from one to the other which in this case is given by   \(\psi(x)= [{\rm Re}(x^{\rm T}),{\rm Im}(x^{\rm T})]^{\rm T}\) where \(x\in\comp^{ {t_{1}}}\).  Let \(\check\psi(x)= [{\rm -Im}(x^{\rm T}),{\rm Rm}(x^{\rm T})]^{\rm T}\) then \({\cal C}\left( B_{q}  \right)\) is mapped into \({\cal V}_q=\text{span} ( \psi(\tilde h^{} _{1} ),\check\psi( \tilde h _{1} ),...,\) \(\psi(\tilde h _{ {r_{2}}} ),\check\psi(h_{ {r _{2}}} ),\psi( h _{1} ),\check\psi( h_{1} ),...,\psi( h _{q-1} ),\check\psi( h_{q-1} ),\psi( h _{q+1} ),\check\psi( h_{q+1} ),...,\psi( h _{ {r_{1}}} ),\check\psi( h_{ {r_{1}}} ) )\).  Thus,
\(
h^{}_{q}\in{\cal C}(B_{q})\) iff  \(\psi(\hvec^{}_{q})\in {\cal V }\) or \(\tilde\psi(\hvec^{11}_{q})\in {\cal V }\)\footnote{or in other words
\(\hvec^{}_{q}\in{\cal C}(
\Hmat_{21})\) iff  \({\cal H}_{\cal V} (\psi(\hvec_{q}))=\psi( \hvec_{ q})\) or  \({\cal H}_{\cal V^{\bot }}(\check \psi(\hvec_{q}))=\check \psi(\hvec_{q}) {\cal  }\) where \({\cal H}_{\cal V}\) is the projection operator into \({\cal V}\).}. Because  \(\tilde \psi(\hvec^{11}_{i})\)   and  \(\psi(\hvec^{11}_{i})\) are orthogonal, \(
h_{q}\in{\cal C}(B_{q})\) is equivalent to \(\psi(h_{q}^{})
\in{\cal V}^{\bot}\) or \(\psi(\hvec_{q }^{})
\in{\cal V}\).
Henceforth the proof is identical to the real case since   \(m^{2 {t_{1}}}({\cal V})=m^{2 {t_{1}}}({\cal V^{\bot}})=0\)  and because \(\Hmat\) is a continuous random matrix.

\bibliographystyle{ieeetr}

\end{document}